\documentclass[aps,prl,superscriptaddress,showpacs,preprintnumbers,10pt,floatfix]{revtex4}
\usepackage{graphicx,color,amsmath}

\newcommand{\be}{\begin{equation}}
\newcommand{\ee}{\end{equation}}
\newcommand{\bea}{\begin{eqnarray}}
\newcommand{\eea}{\end{eqnarray}}

\newcommand{\bfk}{\mbox{\boldmath $k$}}

\def\kt{k_\perp}
\newcommand{\bfp}{\mbox{\boldmath $p$}}

\newcommand{\bfq}{\mbox{\boldmath $q$}}

\newcommand{\bfS}{\mbox{\boldmath $S$}}

\newcommand{\bfz}{\mbox{\boldmath $z$}}
\newcommand{\bfy}{\mbox{\boldmath $y$}}

\newcommand{\bfL}{\mbox{\boldmath $L$}}

\newcommand{\pup}{p^\uparrow}

\def\avk{\langle k_\perp ^2\rangle}

\def\avkS{\langle k_S ^2\rangle}

\def\lsim{\mathrel{\rlap{\lower4pt\hbox{\hskip1pt$\sim$}}\raise1pt\hbox{$<$}}}
\def\gsim{\mathrel{\rlap{\lower4pt\hbox{\hskip1pt$\sim$}}\raise1pt\hbox{$>$}}}
\def\nostrocostruttino#1\over#2{\mathrel{\mathop{\kern 0pt \rlap
{\hbox{$#1$}}} \hbox{\kern-.135em $#2$}}}

\begin{document}

\title{The Sivers asymmetry in $J/\Psi$ and lepton pair production at COMPASS}

\author{M.~Anselmino}
\affiliation{Dipartimento di Fisica, Universit\`a di Torino,
             Via P.~Giuria 1, I-10125 Torino, Italy}
\affiliation{INFN, Sezione di Torino, Via P.~Giuria 1, I-10125 Torino, Italy}
\author{V.~Barone}
\affiliation{Di.S.I.T., Universit\`a del Piemonte Orientale,
             Viale T. Michel 11, I-15121 Alessandria, Italy}
\affiliation{INFN, Sezione di Torino, Via P.~Giuria 1, I-10125 Torino, Italy}
\author{M.~Boglione}
\affiliation{Dipartimento di Fisica, Universit\`a di Torino,
             Via P.~Giuria 1, I-10125 Torino, Italy}
\affiliation{INFN, Sezione di Torino, Via P.~Giuria 1, I-10125 Torino, Italy}
\date{\today}

\begin{abstract}
The abundant production of lepton pairs via $J/\Psi$ creation at COMPASS,
$\pi^\pm \, \pup \to J/\Psi \, X \to \ell^+ \ell^- X$, allows a measurement 
of the transverse Single Spin Asymmetry generated by the Sivers effect.
The crucial issue of the sign change of the Sivers function in lepton pair
production, with respect to Semi Inclusive Deep Inelastic Scattering 
processes, can be solved. Predictions for the expected magnitude of the 
Single Spin Asymmetry, which turns out to be large, are given.       
       
\end{abstract}


\maketitle

The distribution, in momentum space, of unpolarized quarks and gluons 
inside a transversely polarized nucleon, first introduced by 
Sivers~\cite{Sivers:1989cc,Sivers:1990fh}, 
is one of the eight leading-twist Transverse Momentum Dependent Partonic 
Distribution Functions (TMD-PDFs), which can be accessed through experiments 
and encode our information on the 3-Dimensional nucleon structure. 
The Sivers distribution for unpolarized quarks (or gluons) with
transverse momentum $\bfk_\perp$ inside a proton with 3-momentum $\bfp$ 
and spin $\bfS$, is defined as
\bea
\hat f_ {q/\pup} (x,\bfk_\perp) &=& f_ {q/p} (x,\kt) +
\frac{1}{2} \, \Delta^N \! f_ {q/\pup}(x,\kt)  \;
{\bfS} \cdot (\hat {\bfp}  \times
\hat{\bfk}_\perp) \label{sivnoi} \\
&=& f_ {q/p} (x,\kt) - \frac{k_\perp}{m_p} \>
f_{1T}^{\perp q}(x, k_\perp) \;
{\bfS} \cdot (\hat {\bfp}  \times \hat{\bfk}_{\perp}) \>,
\eea
where $f_{q/p}(x,\kt)$ is the unpolarized TMD-PDF and 
$\Delta^N \! f_ {q/\pup} = (-2 k_\perp/m_p) f_{1T}^{\perp q}$ is 
the Sivers function. 

The Sivers distribution is one of the best known polarized TMD-PDFs and 
has a clear experimental signature~\cite{Airapetian:2009ae, Adolph:2012sp}. 
It is of particular interest for several reasons; one expects it to be related 
to fundamental intrinsic features of the nucleon and to basic QCD properties. 
In fact, the Sivers distribution relates the motion of unpolarized quarks 
and gluons to the nucleon spin $\bfS$; then, in order to build a scalar, 
parity invariant quantity, $\bfS$ must couple to the only other available 
pseudo-vector, that is the parton orbital angular momentum, $\bfL_q$ or 
$\bfL_g$. Another peculiar feature of the Sivers distribution is that its 
origin at partonic level can be traced in QCD interactions between the quarks 
(or gluons) active in inelastic high energy interactions and the nucleon 
remnants~\cite{Brodsky:2002cx,Brodsky:2002rv}; thus, it is expected to be 
process dependent and have opposite sign in Semi Inclusive Deep Inelastic
Scattering (SIDIS) and Drell-Yan (D-Y) 
processes~\cite{Collins:2002kn,Brodsky:2013oya}. This important prediction 
remains to be tested.

The Sivers distribution can be accessed through the study of azimuthal 
asymmetries in polarized SIDIS and D-Y processes. These have been clearly 
observed in the last years, in SIDIS, by the HERMES~\cite{Airapetian:2009ae} 
and COMPASS~\cite{Adolph:2012sp} Collaborations, allowing the first extraction 
of the SIDIS Sivers function~\cite{Anselmino:2008sga,Bacchetta:2011gx,
Anselmino:2012aa}. However, no information could be obtained on the D-Y Sivers 
function, as no polarized D-Y process had ever been measured. 

Asymmetries related to the Sivers effect can also be measured in the so called
generalized D-Y processes~\cite{Peng:2014hta,Huang:2015vpy}, that is the creation 
of lepton pairs via vector bosons, $p \, p \to W^\pm X \to \ell^\pm \, \nu \, X$ 
and $p \, p \to Z^0 \, X \to \ell^+ \ell^- X$. Also in this case one expects a 
Sivers function opposite to that observed in SIDIS.  

Recently, first few data from D-Y weak boson production at RHIC, $\pup \, p \to 
W^\pm/Z^0 \, X$, have become available~\cite{Adamczyk:2015gyk}. 
They show some azimuthal asymmetry which hints, with large errors and sizeable
uncertainties, at a sign change between the Sivers function observed in these 
generalised D-Y processes and the SIDIS Sivers function. More data on genuine 
D-Y processes, $\pi^\pm \, \pup \to \gamma^* \, X \to \ell^+ \ell^- X$, are 
expected soon from the COMPASS Collaboration. However, also in this case, due 
to the energy of the COMPASS experiment, $\sqrt s = 18.9$ GeV, and the accepted 
safe region for D-Y events, $M \gsim 4$ GeV/$c^2$, where $M$ is the invariant 
mass of the lepton pair, only a limited number of events, and consequently 
large statistical errors, are expected.    

Following Refs.~\cite{Anselmino:2004ki,Barone:2006ws} 
and~\cite{Peng:2014hta,Huang:2015vpy} we propose here to measure the lepton 
pair production at COMPASS at the peak of the $J/\Psi$ production, where the 
number of events is greatly enhanced. Notice that the spin-parity quantum 
numbers of $J/\Psi$ are the same as for a photon.   

Let us start from the usual D-Y. According to the TMD factorisation scheme, 
the cross section for this process, $ h_1 \, h_2 \to q\,\bar q\,X 
\to \ell^+ \ell^-\,X$, in which one measures the four-momentum $q$ of the 
lepton pair, can be written, at leading order,  
as~\cite{Anselmino:2009st,Kang:2009bp}:
\be
\frac{d \sigma^{h_1 h_2 \to \ell^+\ell^-X}}{dy \, dM^2 \, d^2\bfq_T} =
\hat{\sigma}_0 \sum_{q} e_q^2
\int d^2\bfk_{\perp 1} \, d^2\bfk_{\perp 2} \>
\delta^2(\bfk_{\perp 1} + \bfk_{\perp 2} - \bfq_T) \>
f_{\bar q/h_1}(x_1, k_{\perp 1}) \>
f_{q/h_2}(x_{2}, k_{\perp 2}) \label{DYunp}
\ee
where the $\sum_q$ runs over all relevant quarks and antiquarks and we  
have adopted the usual variables:
\be
q = (q_0, \bfq_T, q_L) \quad\quad q^2 = M^2 \quad\quad
y = \frac 12 \ln \frac{q_0 + q_L}{q_0 - q_L} \quad \quad
s=(p_{1} + p_{2})^2 \>\cdot
\label{var}
\ee 
The $f_{q/h}(x, k_{\perp})$ are the unpolarized TMD-PDFs and 
$e_q^2 \, \hat{\sigma}_0$ is the cross section for the 
$q \, \bar q \to \ell^+ \ell^-$ process: 
\be
e_q^2\, \hat{\sigma}_0 = e_q^2 \, \frac{4 \pi \alpha^2}{9 M^2} \> \cdot
\label{s0}
\ee  
$\bfk_{\perp 1}$ and $\bfk_{\perp 1}$ are the parton transverse momenta, 
while the parton longitudinal momentum fractions are given by
\be
x_{1,2} = \frac{M}{\sqrt s} \, e^{\pm y}, \quad {\rm so \> that} \quad
x_F = \frac{2\,q_L}{\sqrt s} = x_1 - x_2 
= \left( x_1 - \frac {M^2}{s \,x_1} \right)
= \left( \frac {M^2}{s \,x_2} - x_2 \right),
\quad y = \frac 12 \ln \frac {x_1}{x_2} = 
\ln \frac {x_1\, \sqrt s}{M} \> \cdot  \label{x1x2}
\ee
Eq.~(\ref{DYunp}) holds in the kinematical region: 
\be
q_T^2 \ll M^2 \quad\quad\quad k_{\perp} \simeq q_T \>. \label{kinr}
\ee

In the case in which one of the hadrons, say $h_2^\uparrow$, is polarized, 
Eq.~(\ref{DYunp}) simply modifies by replacing $f_{q/h_2}(x_2, k_{\perp 2})$
with $\hat f_{q/h_2^\uparrow} (x_2,\bfk_{\perp 2})$ as given in 
Eq.~(\ref{sivnoi}). We then have the Sivers single transverse spin asymmetry:
\bea
A_N &=& \frac{d\sigma^{h_1 h_2^\uparrow \to \ell^+ \ell^- X}
          - d\sigma^{h_1 h_2^\downarrow \to \ell^+ \ell^- X}}
           {d\sigma^{h_1 h_2^\uparrow \to \ell^+ \ell^- X}
          + d\sigma^{h_1 h_2^\downarrow \to \ell^+ \ell^- X}}
\equiv \frac{d\sigma^\uparrow - d\sigma^\downarrow}
           {d\sigma^\uparrow + d\sigma^\downarrow} \label{asy} \\
&=& \frac
{\sum_{q} e_q^2 
\int d^2\bfk_{\perp 1} \, d^2\bfk_{\perp 2} \>
\delta^2(\bfk_{\perp 1} + \bfk_{\perp 2} - \bfq_T) \>
{\bfS} \cdot (\hat {\bfp}_2  \times \hat{\bfk}_{\perp 2}) \>
f_{\bar q/h_1}(x_{1}, k_{\perp 1}) \>
\Delta^N\!f_{q/h_2^\uparrow}(x_2, k_{\perp 2})}
{2 \sum_{q} e_q^2
\int d^2\bfk_{\perp 1} \, d^2\bfk_{\perp 2} \>
\delta^2(\bfk_{\perp 1} + \bfk_{\perp 2} - \bfq_T) \>
f_{\bar q/h_1}(x_1, k_{\perp 1}) \>
f_{q/h_2}(x_{2}, k_{\perp 2})} \label{ann} \> \cdot
\eea

When the lepton pair production occurs via $q \, \bar q$ annihilation into a  
vector meson $V$ rather than a virtual photon $\gamma^*$, Eqs.~(\ref{DYunp}), 
(\ref{s0}) and~(\ref{ann}) still hold, with the replacements~\cite{Anselmino:2004ki}:
\be
16 \pi^2 \alpha^2 e_q^2 \to (g_q^V)^2 \, (g_\ell^V)^2 \quad\quad\quad
\frac 1{M^4} \to \frac 1{(M^2 - M_V^2)^2 + M_V^2\,\Gamma_V^2} \>,
\ee
where $g_q^V$ and $g_\ell^V$ are the $V$ vector couplings to $q \, \bar q$ and
$\ell^+ \ell^-$ respectively. $\Gamma_V$ is the width of the vector meson and 
the new propagator is responsible for a large increase in the cross section at 
$M^2 = M_V^2$. 

We then have:
\be
A_N^V = 
\frac{\sum_{q} (g_q^V)^2 
\int d^2\bfk_{\perp 1} \, d^2\bfk_{\perp 2} \>
\delta^2(\bfk_{\perp 1} + \bfk_{\perp 2} - \bfq_T) \>
{\bfS} \cdot (\hat {\bfp}_2  \times \hat{\bfk}_{\perp 2}) \>
f_{\bar q/h_1}(x_{1}, k_{\perp 1}) \>
\Delta^N\!f_{q/h_2^\uparrow}(x_2, k_{\perp 2})}
{2 \sum_{q} (g_q^V)^2
\int d^2\bfk_{\perp 1} \, d^2\bfk_{\perp 2} \>
\delta^2(\bfk_{\perp 1} + \bfk_{\perp 2} - \bfq_T) \>
f_{\bar q/h_1}(x_1, k_{\perp 1}) \>
f_{q/h_2}(x_{2}, k_{\perp 2})} \label{annV} \> \cdot
\ee  
We propose to use Eq.~(\ref{annV}) for lepton pair production at COMPASS,
$\pi^\pm \, \pup \to \ell^+ \ell^- X$, at the $J/\Psi$ peak, $M^2 = M^2_{J/\Psi}$.
There are several reasons which make this channel very interesting and promising.

\vskip 6pt \noindent
{\it 1)} At COMPASS energy one has $x_1\,x_2 = M_{J/\Psi}^2/s \simeq 0.027$.
Due to this relation both $x_1$ and $x_2$ must be greater than 0.027 and one
of them must be greater than $\sqrt{0.027} \simeq 0.16$. At small values 
of $x_F$ or $y$ one has approximately $x_1 \simeq x_2 \simeq 0.16$. It is then 
reasonable to expect that the main channel for the $J/\Psi$ production is 
indeed $q\,\bar q$ annihilation (rather than gluon fusion). 

\vskip 6pt \noindent
{\it 2)}
The COMPASS data which have been taken in 2015 and are being analysed refer 
to the $\pi^- \, \pup \to \ell^+ \ell^- X$ process at $\sqrt s = 18.9$ GeV. 
Their interesting feature is that the dominant contribution to the 
asymmetry~(\ref{annV}) is given by a $\bar u$ quark from the $\pi^-$ and 
a $u$ quark from the proton, both of them valence quarks. All other contributions
would always involve a sea quark and, in the central rapidity region, are 
strongly suppressed.     
     
\vskip 6pt \noindent
{\it 3)} Other production mechanisms of $J/\Psi$ might contribute, like gluon 
fusion. However, while they might enhance the unpolarized cross section, 
the denominator of $A_N^V$, it is very unlikely that they significantly
affect the numerator of $A_N^V$; in fact the gluon Sivers function is 
expected to be small, if not zero~\cite{D'Alesio:2015uta}. Thus, such 
contributions might decrease the value of $A_N^V$, but they cannot alter the 
conclusion that it mainly originates from the valence quark Sivers functions.  

Then we have, for central rapidity $\pi^- \, \pup \to J/\Psi \, X \to 
\ell^+ \ell^- X$
processes:
\be
A_N^{J/\Psi}(\pi^-; x_1, x_2, \bfq_T) \simeq 
\frac {\int d^2\bfk_{\perp 1} \, d^2\bfk_{\perp 2} \>
\delta^2(\bfk_{\perp 1} + \bfk_{\perp 2} - \bfq_T) \>
{\bfS} \cdot (\hat {\bfp}_2  \times \hat{\bfk}_{\perp 2}) \>
f_{\bar u/\pi^-}(x_{1}, k_{\perp 1}) \,
\Delta^N\!f_{u/p^\uparrow}(x_2, k_{\perp 2}) \>}
{2 \int d^2\bfk_{\perp 1} \, d^2\bfk_{\perp 2} \>
\delta^2(\bfk_{\perp 1} + \bfk_{\perp 2} - \bfq_T) \>
f_{\bar u/\pi^-}(x_1, k_{\perp 1}) \>
f_{u/p}(x_{2}, k_{\perp 2})} \label{annpi-}
\ee 
and, for $\pi^+ \, \pup \to J/\Psi \, X \to \ell^+ \ell^- X$ processes:
\be
A_N^{J/\Psi}(\pi^+; x_1, x_2, \bfq_T) \simeq 
\frac {\int d^2\bfk_{\perp 1} \, d^2\bfk_{\perp 2} \>
\delta^2(\bfk_{\perp 1} + \bfk_{\perp 2} - \bfq_T) \>
{\bfS} \cdot (\hat {\bfp}_2  \times \hat{\bfk}_{\perp 2}) \>
f_{\bar d/\pi^+}(x_{1}, k_{\perp 1}) \,
\Delta^N\!f_{d/p^\uparrow}(x_2, k_{\perp 2}) \>}
{2 \int d^2\bfk_{\perp 1} \, d^2\bfk_{\perp 2} \>
\delta^2(\bfk_{\perp 1} + \bfk_{\perp 2} - \bfq_T) \>
f_{\bar d/\pi^+}(x_1, k_{\perp 1}) \>
f_{d/p}(x_{2}, k_{\perp 2})} \label{annpi+} \> \cdot
\ee 
Notice that the variables $x_1$ and $x_2$ are related to each other and one 
can use only one of them or the variable $x_F$ or $y$, Eq.~(\ref{x1x2}) with 
$M^2 = M_{J/\Psi}^2$. 

Eqs.~(\ref{annpi-}) and (\ref{annpi+}) can be further evaluated, adopting, 
as usual, a Gaussian factorized form both for the unpolarized distribution 
and the Sivers functions, as in Ref.~\cite{Anselmino:2008sga}:
\bea
f_{q/p}(x,k_\perp) &=& f_q(x) \, \frac{1}{\pi \langle\kt^2\rangle} \,
e^{-{\kt^2}/{\langle\kt^2\rangle}} \label{partond}\\
\Delta^N \! f_{q/\pup}(x,\kt) &=& 2 \, {\cal N}_q(x) \, 
h(\kt) \, f_ {q/p} (x,\kt) \label{sivfac} \\
h(\kt) &=& \sqrt{2e}\,\frac{k_\perp}{M_{1}}\,e^{-{k_\perp^2}/{M_{1}^2}} \>,
\label{siverskt}
\eea
where the $f_q(x)$ are the unpolarized PDFs, $M_1$ is a parameter which allows 
the $k_\perp$ Gaussian dependence of the Sivers function to be different from 
that of the unpolarized TMDs and ${\cal N}_q(x)$ is a function which 
parameterises the factorized $x$ dependence of the Sivers function.    
In such a case the $\bfk_\perp$ integrations can be performed analytically in
Eqs.~(\ref{annpi-}) and (\ref{annpi+}), obtaining:
\bea
\hspace*{-0.5cm}
A_N^{J/\Psi}(\pi^-; x_2, \bfq_T) &=& 
\, \frac{\avkS^2}{[\avkS + \avk]^2} 
\> \exp\Biggl[ - \frac{q_T^2}{2\,\avk} \, \Biggl(\frac{\avk - \avkS} 
{\avk + \avkS} \Biggr) \Biggr] 
\frac{\sqrt{2\,e}\,q_T}{M_1} \times
2\,{\cal N}_{u}(x_2) \>
{\bfS} \cdot (\hat {\bfp}_2  \times {\hat\bfq}_{T}) \label{Apm1}\\
&\equiv& A_N^{J/\Psi}(\pi^-; x_2, q_T) \> 
{\bfS} \cdot (\hat {\bfp}_2  \times {\hat\bfq}_{T}) \label{Apm2}
\eea
and
\bea
\hspace*{-0.5cm}
A_N^{J/\Psi}(\pi^+; x_2, \bfq_T) &=& 
\, \frac{\avkS^2}{[\avkS + \avk]^2} 
\> \exp\Biggl[ - \frac{q_T^2}{2\,\avk} \, \Biggl(\frac{\avk - \avkS} 
{\avk + \avkS} \Biggr) \Biggr] 
\frac{\sqrt{2\,e}\,q_T}{M_1} \times
2\,{\cal N}_{d}(x_2) \>
{\bfS} \cdot (\hat {\bfp}_2  \times {\hat\bfq}_{T}) \label{App1}\\
&\equiv& A_N^{J/\Psi}(\pi^+; x_2, q_T) \> 
{\bfS} \cdot (\hat {\bfp}_2  \times {\hat\bfq}_{T}) \label{App2}
\eea
where
\be
\avkS = \frac{ M_1^2 \, \langle k_{\perp}^2\rangle}
{M_1^2 + \langle k_{\perp}^2\rangle} \>\cdot
\ee

$A_N^{J/\Psi}(\pi^\pm; x_2, q_T)$ is the amplitude of the azimuthal modulation 
in the angle
defined by ${\bfS} \cdot (\hat {\bfp}_2 \times {\hat\bfq}_{T})$. For example,
taking the proton moving in the $-\hat\bfz$ direction and $\bfS \equiv \, 
\uparrow$ along $+\hat\bfy$, in the $\pi-p$ c.m. frame, one has 
${\bfS} \cdot (\hat {\bfp}_2  \times {\hat\bfq}_{T}) = - \cos\phi$, where 
$\phi$ is the azimuthal angle of the $J/\Psi$.    

Measurements of $A_N^{J/\Psi}(\pi^-; x_2, q_T)$ and $A_N^{J/\Psi}(\pi^+; x_2, q_T)$ 
give a direct access, respectively, to ${\cal N}_{u}(x_2)$ and ${\cal N}_{d}(x_2)$, 
and the corresponding Sivers functions, Eq.~(\ref{sivfac}). We conclude with an 
estimate of these two quantities based on the Sivers functions extracted from 
SIDIS data. All quantities necessary to compute $A_N^{J/\Psi}(\pi^-; x_2, q_T)$ 
and $A_N^{J/\Psi}(\pi^+; x_2, q_T)$ can be found in Ref.~\cite{Anselmino:2012aa} 
(Eq.~(40) and third column of Table III), taking into account only the valence quark 
contributions. As the Sivers effect is expected to be process dependent and
contribute with different signs to asymmetries in D-Y and SIDIS processes, 
the Sivers functions of Ref.~\cite{Anselmino:2012aa} are used here {\it with 
an opposite sign}.    

\begin{figure}[t]
\begin{center}
\includegraphics[width=7.5truecm,angle=0]{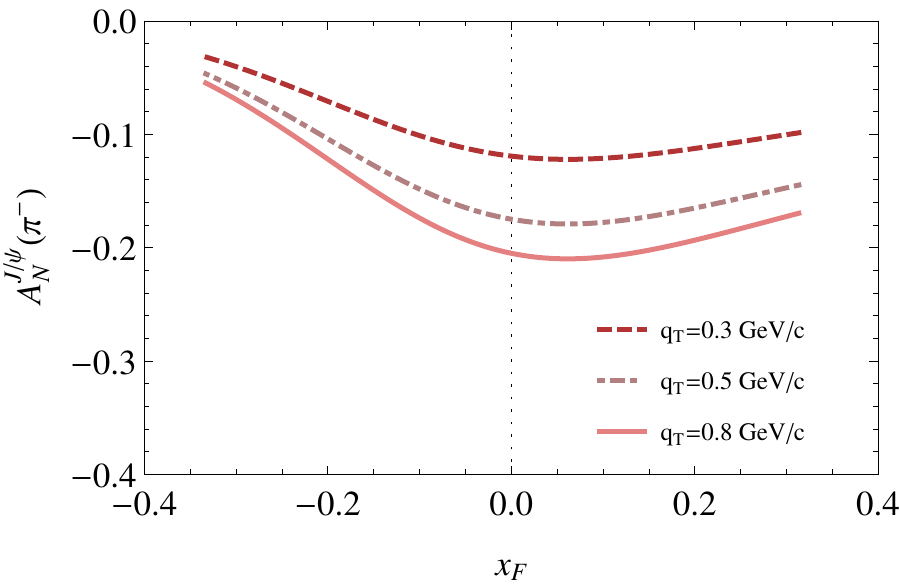}
\hskip 36pt
\includegraphics[width=7.5truecm,angle=0]{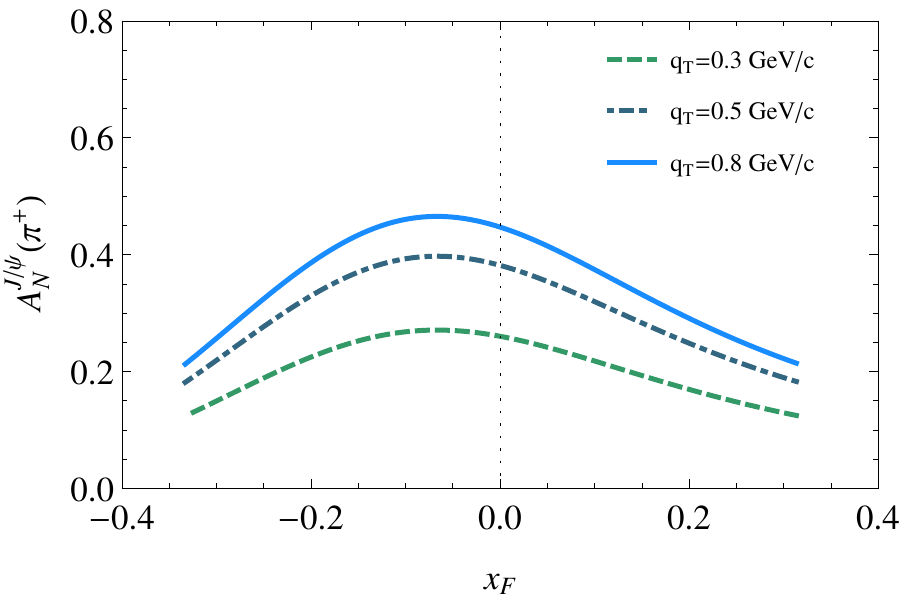}
\caption{\small Plots of $A_N^{J/\Psi}(\pi^-; x_2, q_T)$ (left) and 
$A_N^{J/\Psi}(\pi^+; x_2, q_T)$ 
(right) versus $x_F$, for three different values of $q_T$. These estimates are
obtained according to Eqs.~(\ref{Apm1})--(\ref{App2}) of the text, using the 
parameters of Ref.~\cite{Anselmino:2012aa}, with a sign change for the Sivers 
functions.}
\label{fig1}
\end{center}
\end{figure}

\begin{figure}[t]
\begin{center}
\includegraphics[width=7.5truecm,angle=0]{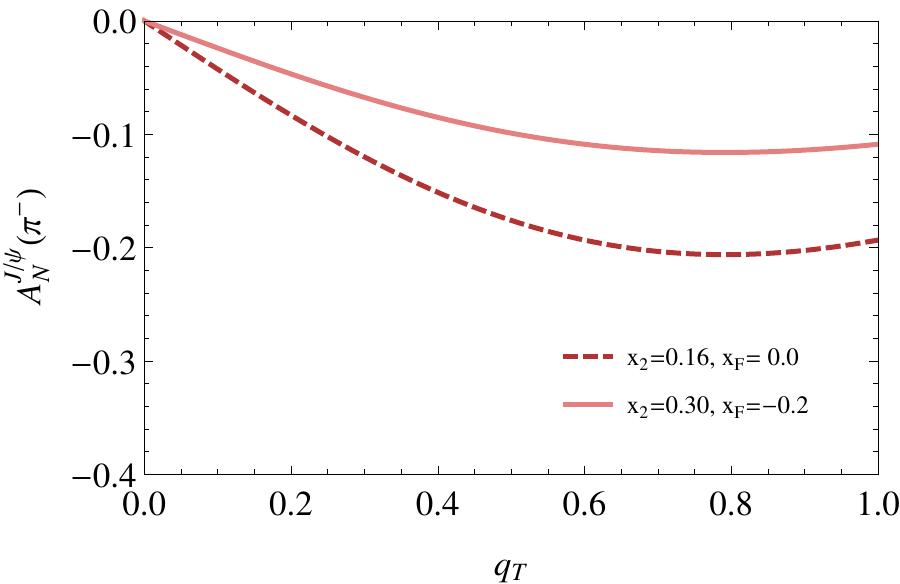}
\hskip 36pt
\includegraphics[width=7.5truecm,angle=0]{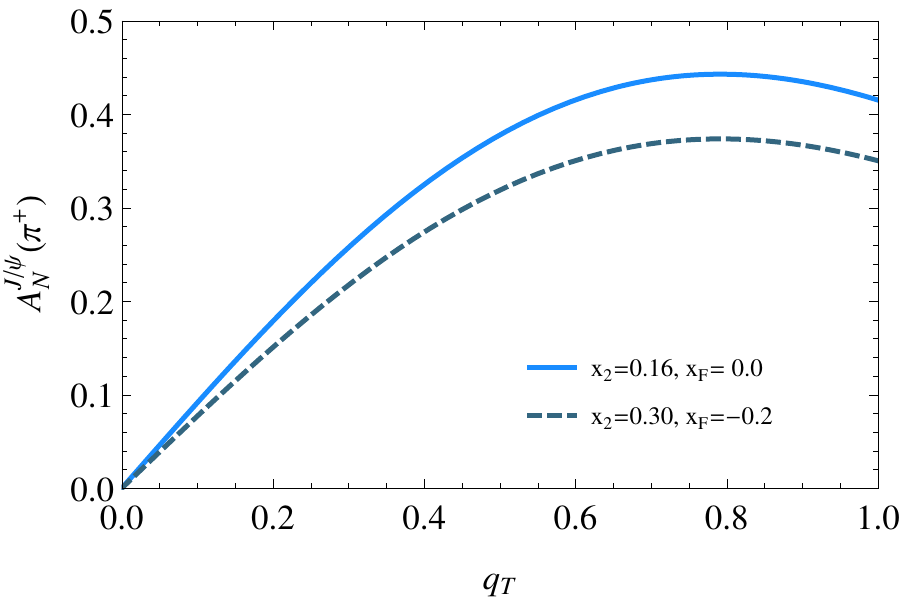}
\caption{\small Plots of $A_N^{J/\Psi}(\pi^-; x_2, q_T)$ (left) and 
$A_N^{J/\Psi}(\pi^+; x_2, q_T)$ 
(right) versus $q_T$, for two different values of $x_F$. These estimates are
obtained according to Eqs.~(\ref{Apm1})--(\ref{App2}) of the text, using the 
parameters of Ref.~\cite{Anselmino:2012aa}, with a sign change for the Sivers 
functions. }
\label{fig2}
\end{center}
\end{figure}

In Fig.~\ref{fig1} we plot $A_N^{J/\Psi}(\pi^-; x_2, q_T)$ (left plot) and 
$A_N^{J/\Psi}(\pi^+; x_2, q_T)$ (right plot), for different values of $q_T$, as 
functions of $x_F$ in the expected kinematical region of the COMPASS experiment.
Similarly, in Fig.~\ref{fig2} we plot the asymmetries, for different values of 
$x_F$, versus $q_T$. 

In both cases the Sivers asymmetries are large, with a well defined sign, driven
by the sign of the Sivers functions of the proton valence quarks, $u$ quark for 
$A_N^{J/\Psi}(\pi^-)$ and $d$ quark for $A_N^{J/\Psi}(\pi^+)$. We consider these 
large values as a definite indication of the sign of the Sivers functions. 
Taking into account the uncertainty bands of the Sivers functions in 
Ref.~\cite{Anselmino:2012aa} would change the expected magnitudes of 
$A_N^{J/\Psi}(\pi^-)$ and $A_N^{J/\Psi}(\pi^+)$, but not
their signs. Notice that, in order to obtain better statistics, one could 
gather data over the full range of $q_T$ for which Eq.~(\ref{kinr}) holds; 
then the asymmetries are given by Eqs.~(\ref{annpi-}) and~(\ref{annpi+}) with 
numerator and denominator integrated over $q_T$ from 0 to, say, 1 GeV/$c$. 

In conclusion, we propose a simple measurement of the single transverse spin 
asymmetry $A_N$ in the channel $\pi^\pm \, \pup \to J/\Psi \, X \to 
\ell^+ \ell^- X$, for which abundant data have been already collected by the 
COMPASS Collaboration. Due to the kinematical feature of the experiment, 
the asymmetry is mainly generated by the Sivers distribution of unpolarized
valence quarks inside the polarized proton and its sign reveals the sign of 
the corresponding Sivers function. Thus, the longstanding debate about the 
opposite sign of the Sivers function in SIDIS and Drell-Yan processes can 
be unambiguously solved.   

\vskip 12pt
\noindent
M.A. and M.B.~acknowledge support from the ``Progetto di Ricerca Ateneo/CSP" 
(codice TO-Call3-2012-0103). 

\bibliography{sample}


\end{document}